\documentclass[letterpaper,twocolumn,10pt]{article}
\usepackage{usenix,epsfig,endnotes}

\usepackage{pdflscape}
\usepackage{xcolor}
\usepackage{float} 
\usepackage{amsmath}
\usepackage{makecell} 
\usepackage{comment}
\makeatletter

\renewcommand{\textcolor}[2]{#2}
\makeatother

\usepackage{float}

\usepackage[english]{babel}
\usepackage{graphicx}
\usepackage{graphics}
\usepackage[utf8]{inputenc}
\usepackage{amsmath}
\usepackage{tabularx}
\usepackage{array, multirow} 
\usepackage{makecell}
\usepackage{enumitem}
\usepackage{array,multirow}
\usepackage[final]{listings}
\usepackage{color}
\usepackage{dblfloatfix} 
\usepackage{etoolbox}
\usepackage{float}
\usepackage{microtype}
\usepackage{subcaption}
\usepackage{blindtext}
\usepackage{filecontents}
\usepackage{multirow}
\usepackage{booktabs}
\usepackage{siunitx}
\usepackage{pifont}
\usepackage{xspace}
\usepackage{amsmath}
\usepackage{bm}
\usepackage{etoolbox}
\usepackage{pifont}
\usepackage{arydshln}
\usepackage{bm}
\usepackage{xurl}
\usepackage{amssymb}

\PassOptionsToPackage{hyphens}{url}\usepackage{hyperref}

\newcommand{\toolname}{\textsc{TurboTest}\xspace}

\def \showcomments {} 
\newcommand{\tarun}[1]{\ifdef{\showcomments}{[\textcolor{blue}{\textit{Tarun: #1}}]}{}}

\newcommand{\smartparagraph}[1]{\noindent{\bf #1}\ }

\setlist[itemize]{left=0pt, itemsep=0pt}
\setlist[enumerate]{left=0pt, itemsep=0pt}

\usepackage[all=normal,wordspacing]{savetrees}

\begin{document}
\title{
\textsc{TurboTest}: \\
Learning When Less is Enough through Early Termination of Internet Speed Tests
}

\author{
{\rm Haarika Manda}\\
UC Santa Barbara
\and
{\rm Manshi Sagar}\\
IIT Delhi
\and
{\rm Yogesh}\\
 IIT Delhi
\and
{\rm Kartikay Singh}\\
 IIT Delhi
\and
{\rm Cindy Zhao}\\
UC Santa Barbara
\and
{\rm Tarun Mangla}\\
IIT Delhi
\and
{\rm Phillipa Gill}\\
Google
\and
{\rm Elizabeth Belding}\\
UC Santa Barbara
\and
{\rm Arpit Gupta}\\
UC Santa Barbara
}


\maketitle
\begin{sloppypar}

\begin{abstract}

Internet speed tests are indispensable for users, ISPs, and policymakers, but their static flooding-based design imposes growing costs: a single high-speed test can transfer hundreds of MB, and collectively, platforms like Ookla, M-Lab, and Fast.com generate petabytes of traffic each month. Reducing this burden requires deciding when a test can be stopped early without sacrificing accuracy. We frame this as an optimal stopping problem and show that existing heuristics--static thresholds, BBR pipe-full signals, or throughput stability rules from Fast.com and FastBTS--capture only a narrow slice of the achievable accuracy--savings trade-off. This paper introduces \textsc{TurboTest}, a systematic framework for  speed test termination that sits atop existing platforms. The key idea is to decouple throughput prediction (Stage~1) from test termination (Stage~2): Stage~1 trains a regressor to estimate final throughput from partial measurements, while Stage~2 trains a classifier to decide when sufficient evidence has accumulated to stop. Leveraging richer transport-level features (RTT, retransmissions, congestion window) alongside throughput, \textsc{TurboTest} exposes a single tunable parameter $\epsilon$ for accuracy tolerance and includes a fallback mechanism for high-variability cases. Evaluation on 1 million M-Lab NDT speed tests (2024--2025) shows that \textsc{TurboTest} achieves 1.8-$4.4\times$ higher data savings than an approach based on BBR signals while reducing median error. These results demonstrate that adaptive ML-based termination can deliver accurate, efficient, and deployable speed tests at scale.

\end{abstract}

\section{Introduction}
\label{sec:intro}  

Internet speed tests are critical tools for multiple stakeholders: users rely on them to verify service quality against advertised speeds, Internet Service Providers (ISPs) to plan capacity and diagnose performance issues, and policymakers to inform broadband policy and evaluate infrastructure investments. However, these measurements impose substantial costs that scale with network performance improvements. Their static, flooding-based design requires saturating the network path for a fixed duration (often 10+ seconds), transferring hundreds of megabytes on high-speed links. This creates significant expenses for providers operating measurement infrastructure and for users on metered connections. With millions of tests conducted daily across platforms such as Ookla~\cite{ookla}, M-Lab~\cite{mlabspeedtest}, and Netflix's Fast.com~\cite{fastspeedtest}, the aggregate traffic volume and infrastructure burden represent a substantial and growing economic cost. Reducing measurement overhead is therefore essential, not only for providers but also for preserving the societal value of speed test data. 

Different providers have embraced distinct \emph{speed test methodologies}. For example, Ookla employs multi-threaded tests that offer high fidelity at the cost of substantial overhead, while M-Lab uses a single-threaded design with BBR congestion control to saturate the link. Newer methodologies such as FastBTS~\cite{fastbts} and Fast.com~\cite{fastspeedtest} aim to balance efficiency and accuracy by integrating convergence-based stopping rules. However, scaling new methodologies remains challenging due to legacy constraints, proprietary business models, and lack of standardization. This paper does \emph{not} focus on developing yet another measurement methodology. Instead, it addresses the complementary problem of designing an \emph{external termination layer} that sits atop existing methodologies: given an ongoing test, determine when it can be stopped early without sacrificing accuracy. This layered approach makes solutions deployable across diverse platforms without replacing existing test designs. 

\smartparagraph{The opportunity: early termination.} 
This problem can be framed as an \emph{optimal stopping} problem. At each point in the test, the decision maker must weigh the potential accuracy gain from continuing the test against the savings from stopping early. The challenge is to design stopping policies that balance these competing objectives while remaining robust across heterogeneous and evolving access networks.  

\smartparagraph{Existing approaches.}  
Several heuristic approaches to test termination have been proposed for this problem. Static thresholds---e.g., M-Lab's fixed 250~MB cap~\cite{mlabspeedtest} or Cloudflare's capped tests~\cite{cloudflarespeedtest}---offer simplicity but ignore link heterogeneity. Transport-signal rules, such as BBR's pipe-full indicator~\cite{bbr}, \textcolor{red}{have also been used to exploit congestion-control signals}.
Other ideas were not originally designed for external termination but could, in principle, be applied to it. For example, Netflix's Fast.com implements a throughput stability heuristic (TSH)~\cite{fastspeedtest}, and FastBTS introduced crucial interval sampling (CIS)~\cite{fastbts}; while these are embedded in complete methodologies, the underlying logic could also be repurposed as external stopping rules. Each of these methods defines a stopping condition and exposes a tunable parameter that governs the trade-off between accuracy and efficiency.  

\smartparagraph{Limitations of heuristics.}  
Despite their utility, all such approaches share key limitations. First, they rely on fragile assumptions.  For instance, BBR fails in high-speed links where pipe-full may never appear.
Second, they draw only on narrow signal spaces (e.g. throughput time series), discarding richer TCP-level features such as RTT and congestion window dynamics that could provide more reliable evidence of convergence. Third, even when they terminate at the ``right'' time, their throughput estimation remains naïve and utilizes simple averages that yields biased results. As a result, these approaches capture only a narrow slice of the achievable accuracy--savings frontier.


\smartparagraph{Our approach: \textsc{TurboTest}.}  
This paper introduces \textsc{TurboTest}, a systematic framework for early termination of Internet speed tests. The key idea is to decompose the problem into two coordinated subproblems: (i) \emph{prediction}, where a regression model estimates the final throughput from partial measurements, and (ii) \emph{termination}, where a classifier decides when to stop based on prediction accuracy and cost trade-offs. This decomposition avoids conflating two separate but not independent tasks: termination decision are conditioned on prediction outcomes, accurate predictions rely on termination to trigger only once sufficient evidence has accumulated. This coupling enables more aggressive yet reliable termination compared to prior heuristics.  

\textsc{TurboTest} leverages machine learning (ML) to optimize prediction and termination independently, while coordinating them end-to-end. ML naturally incorporates complementary metrics and adapts to non-stationary network conditions that break heuristic assumptions. It can also exploit the vast repositories of \emph{full} speed test results already collected by providers, turning existing infrastructure into a training resource. To ensure robustness in deployment, \textsc{TurboTest} includes a lightweight fallback mechanism: tests exhibiting high variability---where early termination would be unreliable---are allowed to run to completion, bounding worst-case error.  

\smartparagraph{Key contributions.}  
This paper makes three contributions:  
\begin{itemize}
    \item \textbf{Learning problem.} We cast external termination of speed tests as an \emph{optimal stopping} problem, exposing the compound trade-off between accuracy and measurement cost and clarifying the implicit assumptions behind existing heuristics.  
    \item \textbf{Design.} We develop \textsc{TurboTest}, a two-stage ML framework that decouples throughput prediction from termination, leverages richer transport-level features (e.g., RTT, retransmissions, congestion window) alongside throughput, and exposes a single tunable parameter $\epsilon$ that encodes operator accuracy tolerance.
    
    \item \textbf{Evaluation.} We evaluate \textsc{TurboTest} on \textcolor{red}{1 million M-Lab speed tests} spanning 2024--2025 across diverse access types. Our results demonstrate that \textsc{TurboTest} consistently improves the accuracy-savings frontier over state-of-the-art heuristics. At $\epsilon=15$, for example, \textsc{TurboTest} achieves 91.7\% data savings with 18.6\% median error, outperforming BBR's maximum of 84.9\% savings at \textcolor{red}{35.4\%} error and CIS's \textcolor{red}{89.9\%} data savings at \textcolor{red}{32.1\%} error (see \S\ref{ssec:eval_global}). Importantly, it closes much of the gap to oracle bounds, showing that adaptive ML-based termination can approach theoretical limits while remaining deployable at scale. 
\end{itemize}

The remainder of this paper is organized as follows. Section~\ref{sec:background} reviews Internet speed test methodologies; Section~\ref{sec:motivation} formalizes the external termination problem. Section~\ref{sec:methodology} presents the design of \textsc{TurboTest}, detailing its two-stage framework for prediction and termination as well as implementation specifics. Section~\ref{sec:evaluation} evaluates \textsc{TurboTest} against heuristic baselines, exploring accuracy--savings trade-offs, adaptive parameterization, and robustness. Section~\ref{sec:related} summarizes the related work and Section~\ref{sec:discussion} concludes with future directions.

\section{Early Termination of Speed Tests}
\label{sec:background}
\subsection{Speed Tests in Practice}

Internet speed tests are widely used to evaluate end-to-end performance and broadband quality. At a high level, these tests estimate bottleneck link throughput by saturating one or more TCP connections to strategically placed servers. Different providers adopt different strategies to achieve this goal.
Tests employ different termination strategies to ensure that there is self-congestion at the bottleneck link: M-Lab uses a fixed 10-second duration, whereas Ookla~\cite{ookla} and Fast.com~\cite{fastspeedtest} terminate dynamically.

Despite methodological differences, all approaches consume significant network resources. This problem -- worsened by increasing access speeds -- creates strain on test provider infrastructure and incurs costs for users with data caps. For example, M-Lab reports that its global test infrastructure generated 12~PB of traffic in September 2024 alone, a 23$\times$ increase compared to the same period in 2023~\cite{bbr}. 

\subsection{The Early Termination Problem}
Speed test overhead has been recognized as a problem, with new protocols being developed to balance accuracy with data usage. For instance, FastBTS proposes a modified congestion control approach that  saturates the bottleneck link faster~\cite{fastbts}. 
Recent interest has shifted towards reducing overhead by terminating the test earlier, thus augmenting existing methods rather than replacing them. For example, M-Lab recently capped tests at 250~MB~\cite{mlabDataCaps}. This reduces data usage but can underestimate throughput, particularly on high-speed networks. More broadly, early termination introduces an  accuracy--efficiency tradeoff. Stopping sooner reduces bandwidth consumption and server load but risks underestimating the user's true throughput, while continuing longer increases accuracy but inflates costs. 

This leads to a new decision problem: given partial measurements collected during an ongoing test, can we stop without significantly degrading the accuracy of the reported throughput? An effective early termination solution must balance accuracy and data usage, generalizing across heterogeneous access types (e.g., cable, fiber, cellular) and dynamic network conditions. This motivates the rule-based heuristics we discuss next (\S\ref{sec:terminationapproaches}) and the need for a systematic framework that can push the accuracy--savings frontier outward.

\subsection{Existing Early Termination Approaches}
\label{sec:terminationapproaches}
Several efforts have explored rule-based heuristics for terminating speed tests. These heuristics embody simple decision rules that map partial throughput observations to a stopping decision. We group them into two broad classes: static thresholds and dynamic convergence-based rules.

\smartparagraph{Static thresholds.}  
The simplest approach is to terminate after transferring a fixed amount of data, such as 10~MB, 100~MB, or 1~GB. The rationale is straightforward: larger transfers improve accuracy in estimating bottleneck throughput, while smaller transfers reduce measurement cost. 
However, such thresholds are oblivious to network heterogeneity: a 10~MB transfer may suffice for a 25~Mbps connection but be grossly inadequate for a 1~Gbps link. Thus static thresholds, while attractive for their simplicity, are poor at controlling error across diverse conditions due to their lack of adaptivity~\cite{bbr}.

\smartparagraph{BBR pipe-full termination.}  
A more dynamic heuristic leverages the TCP BBR congestion control algorithm~\cite{bbr}. BBR estimates bottleneck bandwidth by tracking the maximum delivery rate and registering a ``pipe-full'' signal once it detects the link is saturated. This pipe-full signal can be used to stop the test after a specified number of pipe-full events (e.g., 3, 5, or 7). A smaller value terminates more aggressively, saving bandwidth but risking underestimation, while larger values improve accuracy at the cost of higher bandwidth consumption. Although this heuristic adapts to congestion dynamics, it relies on a narrow transport-layer signal and struggles in high-speed tests where pipe-full events occur late or not at all.

\smartparagraph{Throughput stability heuristic (TSH).}  
TSH is a dynamic heuristic originated by Netflix's Fast.com~\cite{fastspeedtest}, which is a complete test design rather than an explicit early termination solution. The key idea is to monitor throughput over time and terminate the test once the throughput remains within a small tolerance or threshold. This method can be adapted for early termination more generally: a test could be stopped once throughput fluctuations fall within the tolerance window. Two parameters govern this tradeoff: the tolerance level and the stability window length. Smaller tolerances and longer windows yield higher accuracy but longer tests, while looser tolerances and shorter windows enable earlier termination with greater error risk. TSH is thus more adaptive than static thresholds, but remains vulnerable to variability from bursts, which may delay termination.

\smartparagraph{Crucial interval sampling (CIS).}  
CIS is a dynamic heuristic originated by the FastBTS methodology~\cite{fastbts}. Its central idea is the notion of a \emph{crucial interval}: a narrow range in which most throughput samples concentrate. As a test stabilizes, consecutive crucial intervals become increasingly similar, and a connection is deemed ``converged'' once their similarity exceeds a threshold. While FastBTS embeds CIS into a new test design, it can also be adapted for early termination: a test can be stopped once the difference between consecutive crucial intervals falls below a tunable bound. The threshold, denoted by $\beta$, acts as knob to tune the accuracy--efficiency tradeoff. Like TSH, however, CIS remains sensitive to short-term variability and relies solely on throughput-based signals.\looseness+1

\section{\textsc{TurboTest}'s Motivation}
\label{sec:motivation}
\smartparagraph{Early termination as optimal stopping.} Consider a speed test that runs for $T$ seconds and produces a sequence of samples ${x_i}$ (a vector of network measurements such as throughput and RTT). Let the true throughput $y_{\text{true}}$ be the throughput estimate computed from the full sequence. The early termination problem is to determine a stopping time $\tau \leq T$ and a prediction function $f$ such that the reported throughput $\hat{y} = f(x_1, \dots, x_\tau)$ approximates $y_{\text{true}}$ while minimizing $\tau$, which in turn minimizes data transfer. This naturally fits the classical optimal stopping framework: at each time step $t$, the algorithm observes the prefix ${x_1, \dots, x_t}$ and decides whether to stop ($\tau = t$) or continue ($\tau > t$). The central objective is to balance accuracy -- ensuring the estimation error, $| \hat{y} - y_{\text{true}} |$, is within $\epsilon$ of the true throughput – and efficiency, i.e., minimizing data transferred before termination.

\smartparagraph{Limitations of existing approaches.}
Heuristic solutions \textcolor{red}{to the early termination problem} are fundamentally limited in three ways. First, they rely on fragile assumptions about network behavior that do not hold uniformly in practice. For example, BBR often fails on very high-throughput tests (e.g., $>$400~Mbps), since the test can complete before enough pipe-full signals are observed~\cite{bbr}.  This is particularly problematic considering that high throughput speed tests transfer a larger amount of data, leading to excess operator costs. CIS  makes assumptions about smooth convergence that can lead to premature termination during transient bursts or wireless variability. 

Second, heuristics exploit only a narrow slice of the signal space, typically the throughput time series or a single transport-layer variable, while ignoring subtler patterns across other fields (e.g., RTT, retransmissions, congestion window). These signals are readily-available through the Linux \texttt{tcp\_info} struct~\cite{tcpinfo} and could provide earlier and more reliable evidence of convergence. 

Third, heuristics conflate two intrinsically distinct learning problems: regression (estimating the final throughput) and classification (deciding when to stop). Even when a heuristic chooses the ``right'' stopping time, naive throughput estimation can yield large errors (e.g., CIS or BBR stopping at the right point but reporting a biased aggregate). 
In reality, regression and classification are tightly coupled: smarter regression reduces error at the stopping point and provides stronger predictive signals for deciding whether to stop.

Taken together, these limitations highlight why existing heuristics struggle to balance accuracy and efficiency across diverse access networks. They either stop too early, stop too late, or misestimate throughput even when they stop at the right time. This motivates the need for a systematic framework that leverages richer signals and explicitly decouples prediction from stopping, as we develop in \toolname.

\smartparagraph{The case for decoupling termination and prediction.} 
The optimal stopping formulation naturally decomposes into two distinct subproblems: \textit{termination} (when to stop) and \textit{prediction} (estimating the final throughput). These subproblems have different objectives and error sensitivities. In particular, prediction is a regression task, whereas termination is a classification task with asymmetric risks – prematurely stopping may be far more costly than stopping a little late. Decoupling the two lets us design predictors and stopping policies independently, improving modularity and extensibility. At the same time, the two tasks are not independent: stopping policies can be made more aggressive by conditioning them on the predictor outputs to enable more aggressive yet accurate termination strategies.


\smartparagraph{The case for machine learning.} 
Early termination is inherently a multi-objective problem, requiring simultaneous trade-offs between accuracy and efficiency that existing single-objective heuristics cannot capture. A machine learning approach naturally encodes these trade-offs in the loss function, giving providers explicit knobs to balance cost against accuracy. Moreover, when trained on large and diverse datasets, ML models can generalize across heterogeneous network conditions without relying on narrow assumptions such as throughput convergence or stationarity. This is particularly feasible here: providers already collect millions of complete tests spanning diverse environments, offering ample training data. Finally, ML can leverage a richer signal space such as those available from \texttt{tcp\_info} structure of the Linux kernel stack. A joint consideration of multiple signals such as RTT and congestion window behavior could provide earlier, more reliable evidence of termination. 


\smartparagraph{Takeaway.} Unlike classic optimal stopping problems, speed tests are non-stationary and highly variable across between access mediums and network conditions. Heuristics that rely on single features or stationary assumptions cannot robustly balance accuracy and efficiency. \toolname addresses this by (i) explicitly decoupling termination from prediction, (ii)~leveraging richer, multi-dimensional features that capture more subtle patterns in transport dynamics, and (iii) exposing tunable parameters so providers can balance accuracy and overhead across a wide range of conditions. By doing so, \toolname is designed to shift the accuracy-savings frontier outward, achieving higher efficiency at any accuracy budget compared to existing heuristics.

\section{\toolname: Design and Implementation}
\label{sec:methodology}

\toolname is a data-driven framework for early termination that leverages  diverse network signals and provides tunable parameters to balance accuracy and efficiency. The key insight is that early termination can be decomposed into two coordinated tasks: (i) a \emph{termination task}, where a classifier $\pi_\phi$ decides whether the current data suffices to stop, and (ii) a \emph{prediction task}, where a regression model $h_\theta$ estimates the final throughput from partial measurements. During inference, the termination task precedes prediction. However, during training, the classifier ($\pi_\phi$) is 
conditioned on the prediction hypothesis ($h_\theta$). This coupling enables aggressive yet accurate termination: regression accuracy sharpens stopping decisions, while classification ensures regression is invoked only when appropriate.

\begin{figure}[t]
    \centering
    \includegraphics[width=1\linewidth]{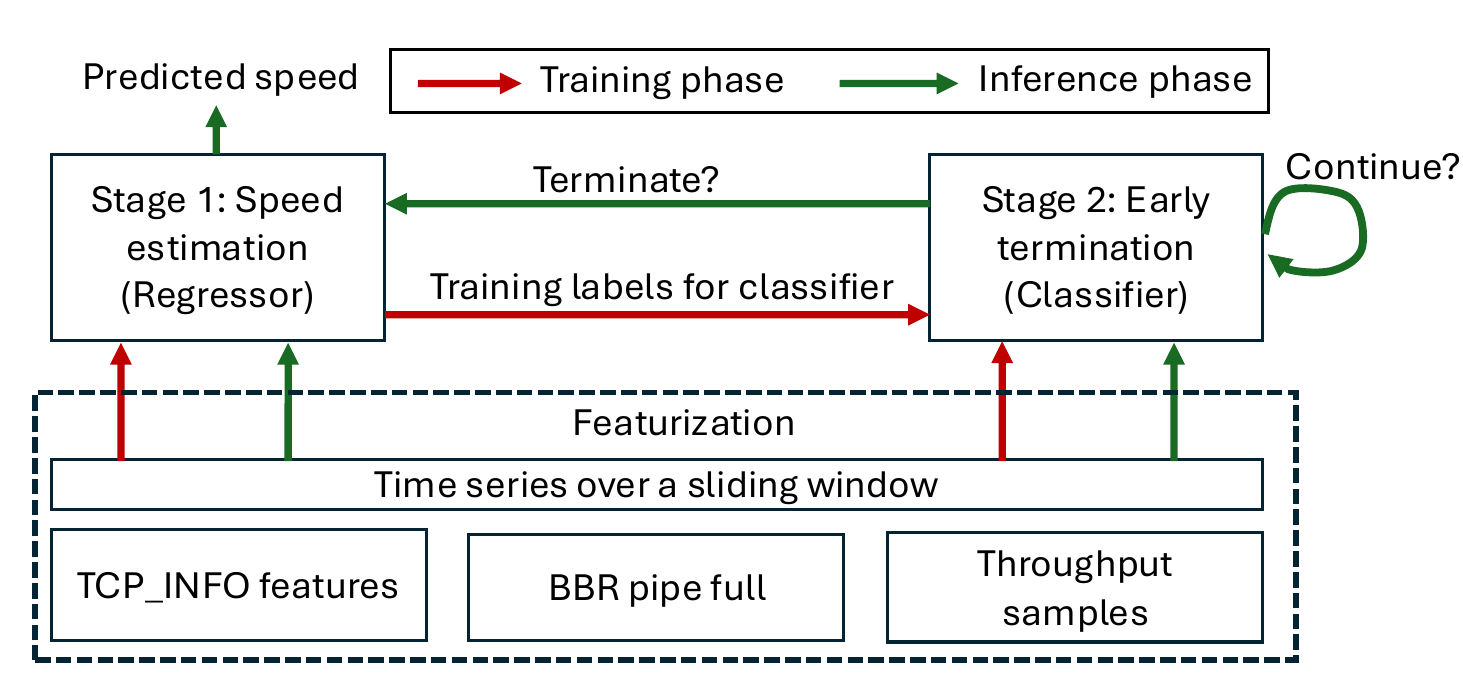}
    \vspace{-1em}
    \caption{\toolname workflow.}
    \label{fig:workflow}
\end{figure}

Figure~\ref{fig:workflow} illustrates the overall workflow. During training, we generate truncated partial sequences from each complete speed test. \textbf{Stage~1} trains a regression model on these sequences; we then compute an \emph{oracle stopping time} $t^*$ -- the earliest point where prediction error falls within tolerance $\epsilon$; and \textbf{Stage~2} trains a stopping classifier to reproduce these oracle decisions, learning when partial data suffices. At inference, the \textbf{Stage~2} classifer operates online as measurements arrive; once it signals termination, the \textbf{Stage~1} regressor predicts the final throughput.

\subsection{Stage 1: Speed Estimation (Regression)}

\smartparagraph{Problem setup.}
The first stage of \textsc{TurboTest} aims to predict the final throughput $y_{\text{true}}$ of a test given only partial observations. This regression task matters because any early termination will be judged by the accuracy of its predicted throughput: even if the stopping decision is correct, poor prediction can still yield large errors. A reliable predictor therefore reduces error regardless of when the test halts and provides the foundation for the stopping policy in Stage~2.  

\smartparagraph{Feature space design.}
A key design question is which features best capture the evolution of a speed test. For instace, because the  TSH, CIS, and BBR-based schemes\footnote{The pipe-full signal is only used for the termination decision.} rely solely on the throughput time series, they miss subtler cues about convergence, particularly under variable conditions (e.g., wireless links or cross traffic). We instead consider three classes of readily available signals: (i) throughput samples, (ii) the pipe-full signal from TCP BBR, and (iii) congestion-control–agnostic transport-layer metrics from the \texttt{tcp\_info} struct in the Linux kernel. The rationale is that richer feature combinations expose patterns not visible to any single signal. 

\smartparagraph{Model architecture.}
We evaluated several models for mapping features to throughput. Linear regression offers interpretability but cannot capture nonlinear dynamics. Tree-based ensembles such as XGBoost handle heterogeneous features, are robust to outliers, and yield interpretable feature importances. Neural networks (feed-forward nets, Transformers) can exploit temporal dependencies. We default to XGBoost for its strong performance on mixed-scale tabular data, fast training, and resilience to missing or sparse inputs -- a critical property for \texttt{tcp\_info} features, where low-throughput tests may send no bytes for hundreds of milliseconds.  We also experimented with neural network architectures, using feed-forward nets as lightweight baselines and Transformers for long-range sequence modeling (see \S\ref{sec:ablation}  for comparison).

\smartparagraph{Training objective.}
Another crucial decision is the choice of training objective. Relative-error losses (e.g., $L_{\text{rel}}(y,\hat y) = \tfrac{|y-\hat y|}{|y|+\gamma}$) emphasize proportional accuracy but can produce unstable gradients as $y \to 0$. In contrast, absolute-error losses such as squared error ($L_{\text{sq}}(y,\hat y) = (y-\hat y)^2$) yield stable optimization and prioritize accuracy at high speeds, leading to greater data savings. Hybrid objectives that combine relative and absolute error are possible but add complexity. We therefore use Mean Squared Error (MSE) for its simplicity and efficiency.\footnote{A limitation is that MSE de-emphasizes relative errors at low speeds.} 
\textcolor{red}{In principle, the regressor could be specialized by throughput regime. For example, one variant could be trained for low-throughput tests (<25 Mbps) and another for high-throughput tests (>400 Mbps). However, for simplicity and ease of deployment, we use a single unified regressor across all tests.}
\subsection{Stage 2: Early Termination (Classification)}

\smartparagraph{Problem setup.}
Stage~2 of \textsc{TurboTest} decides whether sufficient evidence has accumulated to terminate the test at time~$t$. This is naturally framed as a classification problem: given features from the partial sequence, the policy must predict whether additional measurements would materially change the throughput estimate. Unlike Stage~1, which focuses on the \emph{accuracy} of the estimate, Stage~2 focuses on the \emph{sufficiency} of the available information.

\smartparagraph{Label construction.}
To ensure stopping decisions are grounded in prediction quality rather than arbitrary thresholds, we derive oracle labels from Stage~1. For each test $i$, we define the oracle stopping time $t^*_i$ as the earliest point at which the regression prediction error falls within the operator-specified tolerance $\epsilon$. Samples at $t \geq t^*_i$ are labeled as positive (safe to stop), while earlier samples are labeled as negative (must continue). This approach ties classification directly to achievable accuracy, avoiding the brittle parameterization that limits existing heuristics.

\smartparagraph{Feature design.}
A central question is what features the classifier should consume. One option is to base stopping decisions entirely on Stage~1 predictions, mirroring heuristics driven solely by throughput stability. Another is to ingest the same raw features as Stage~1. A third is to collapse prediction and classification into a single joint model. We adopt the second option because it avoids invoking the regressor at every step (lighter than option 1) while preserving modularity between the two tasks.

\smartparagraph{Model architecture.}
We evaluated several alternatives for the classifier, including logistic regression, tree-based ensembles (e.g., XGBoost), and Transformers. We default to Transformers because they empirically yield the best results -- likely because they are designed to capture long-range temporal context and can detect subtle signs of convergence or instability in partial feature sequences.


\smartparagraph{Loss function.}
We train the stopping classifier using standard binary cross-entropy loss. This choice reflects the nature of the task: the model must decide between two outcomes -- whether the current partial sequence is sufficient to stop or not. Binary cross-entropy directly penalizes misclassification, aligns naturally with probabilistic outputs, and provides stable gradients for training. 

\smartparagraph{Training vs. inference.}
An important distinction arises between training and inference. During training, Stage~1 precedes Stage~2: the regression model produces oracle stopping times used as labels for the termination classifier. At inference, however, the order reverses: Stage~2 runs online to decide when to stop the test, after which Stage~1 reports the throughput at the chosen stopping point. This inversion highlights the synergy of the two stages—regression informs classification during training, while classification governs regression at runtime. 

\subsection{\textsc{TurboTest} Implementation}  

We implement \textsc{TurboTest} as a two-stage pipeline consisting of a throughput regressor (Stage~1) and a stopping classifier (Stage~2). Unless otherwise specified, we use an XGBoost regressor in Stage~1 and a Transformer classifier in Stage~2, as validated through ablation studies (\S\ref{sec:ablation}).

\smartparagraph{Features.}  
The feature set includes: (i) throughput samples (instantaneous and cumulative average), (ii) the number of TCP BBR pipe-full signals, and (iii) a subset of metrics from \texttt{tcp\_info}. While \texttt{tcp\_info} provides a rich set of metrics~\cite{tcpinfo}, we focus on congestion window size, bytes in flight, RTT, retransmissions, and duplicate ACKs, as these directly capture throughput, delay, and loss dynamics. NDT already records these metrics at a 10~ms granularity, but we observed that the sampling intervals are not exact and vary across samples. To ensure uniform sequence length and reduce processing cost, we resample these metrics to 100~ms granularity, computing the mean and standard deviation within each window. This yields 13 features per 100~ms interval -- a 10-second test is represented as a 1300-dimensional feature vector. 

\smartparagraph{Model parameters.} We now describe the model parameters of the regressor and classifier:
\begin{itemize}
    \item \textit{Stage~1 (Regression).} We employ XGBoost with depth~7, 1,500 trees, and a learning rate of~0.03, trained using mean squared error (MSE).
    Accounting for our sliding-window training technique, the dataset includes nearly 15 million samples, whose diversity helps mitigate overfitting to specific speed tests.
    \item \textit{Stage~2 (Classification).} We use a transformer model with 8 layers, hidden dimension~128, 8 attention heads, and dropout~0.1, trained with binary cross-entropy loss, the Adam optimizer, learning rate~$10^{-3}$, and batch size~4,096. We intentionally keep the transformer comparatively lightweight to enable fast inference in deployment scenarios. 
\end{itemize}

\smartparagraph{Partial sequence construction.} For model training and inference, we extract partial sequences from the test feature vector. While features are extracted at 100~ms granularity, \textsc{TurboTest} makes termination (and prediction) decisions at 500~ms strides. This design allows us to amortize the model inference overheads, and ensures real-time feasibility. 
For the transformer-based classifier, at time $t$, we use the entire feature history up to $t$. In contrast, the XGBoost-based regressor considers only the most recent two seconds. This is because excessive padding, especially in the early phases of a test, tends to confuse the model, whereas a two second window provides reasonable temporal context. 
For $t < 2$~seconds, we pad the feature vector by duplicating features from the latest 100~ms window, thereby reducing noise from input sparsity.

\smartparagraph{Deployment parameter ($\epsilon$).}  
The only operator-facing parameter is $\epsilon$, which specifies the acceptable error tolerance and is used to define the ground-truth labels for the classifier. For example, with $\epsilon = 20$, the pre-trained regression model identifies the earliest stopping point that keeps the prediction error within 20\%. All subsequent points are labeled as \textit{terminate}, while earlier points are labeled as \textit{continue}. We evaluate across $\epsilon \in \{5, 10, 15, 20, 25, 30, 35\}$, and later extend to an RTT-adaptive setting where $\epsilon$ varies at runtime.

\smartparagraph{Inference workflow.}  
At runtime, each new measurement window is encoded into features and passed to Stage~2. If the classifier outputs \emph{continue}, the test proceeds to the next window. If it outputs \emph{stop}, the regressor is invoked to produce the final throughput estimate, which is returned as the test result. This preserves the two-stage coupling while ensuring that regression is executed only once per terminated test.

 \section{Evaluation}
\label{sec:evaluation}

We 
structure our analysis around five operator-relevant questions: (1) Does \textsc{TurboTest} consistently extend the accuracy--savings Pareto frontier compared to heuristics such as BBR and CIS (Section~\ref{ssec:eval_global})? (2)~Which classes of tests -- defined jointly by throughput tier and RTT -- benefit most from an ML-based termination strategy (Section~\ref{ssec:deepdive})? (3) How can adaptive parameterization, tuned by grouping strategies such as RTT- or speed-based bins, further improve the balance between accuracy and efficiency (Section~\ref{sec:adaptive})? (4) Which design components of \textsc{TurboTest} are most responsible for its performance gains (Section~\ref{sec:ablation})? and (5) How well does \textsc{TurboTest} generalize and scale in practice (Section~\ref{ssec:robustness-overheads})?

\subsection{Experimental Setup}
\smartparagraph{Datasets.}
For evaluation, we use approximately 1~million download speed tests from M-Lab's global measurement platform between April 2024 and March 2025.\footnote{While we focus on NDT due to its publicly available test data, our methodology naturally extends to other tests.} To avoid skew from M-Lab's early termination policy, we exclude all tests pre-terminated by the platform's 250~MB data transfer cap\footnote{Currently, M-Lab randomly applies the data cap on a fraction of their tests.}. 
The data is split into three disjoint subsets: (i)~\emph{training set} of 800k tests (Apr 2024--Jan 2025), balanced across speed tiers. Tiers are defined using thresholds at [25, 100, 200, 400]~Mbps, aligned with policy definitions in the US where links below 25~Mbps and 100~Mbps are classified as unserved and underserved, respectively.  A balanced sampling ensures adequate representation of $>$400~Mbps links, which are fewer but dominate bandwidth overhead. (ii)~\emph{test set} comprising \textcolor{red}{over 1 million} tests (July 2024–Jan 2025), randomly sampled from the natural distribution of speed tests and used for main evaluation. (iii)~\emph{robustness set} of 133k tests (Feb--Mar 2025), used to assess robustness to concept drifts. The random sampling for the latter two datasets preserves the natural distribution of speeds and RTTs in the dataset. \textcolor{red}{Furthermore, random sampling ensures a geographically diverse set of speed tests, spanning 46 countries across 6 continents}.

\begin{figure}[t]
    \centering
    \includegraphics[width=.9\linewidth]{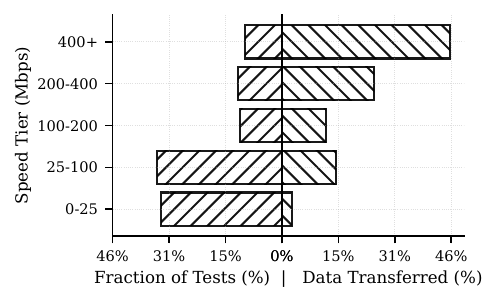}
    \caption{
    Distribution of tests across different speed tiers. The left bars show the fraction of total tests in each group, while the right bars show the fraction of total data transferred.
}
    \label{fig:distribution_skew}
\end{figure}

Figure~\ref{fig:distribution_skew} shows the breakdown of tests across speed tiers, reporting both the fraction of data points and their cumulative contribution to total bandwidth overhead. The imbalance highlights that higher speed tiers, despite having fewer tests, contribute disproportionately to overall bandwidth consumption. As an example, the 400+~Mbps tier includes roughly $4\times$ fewer tests than the 0--25~Mbps tier, yet it contributes about $10\times$ more traffic volume, highlighting the outsized role of high-speed tiers in driving measurement costs. Our splitting strategy ensures the model learns from a balanced training distribution while evaluations reflect real-world conditions, avoiding contamination between training and testing.

\smartparagraph{Baselines.}
We compare \textsc{TurboTest} against three classes of dynamic early termination strategies, each exposing a tunable parameter to balance relative error against data savings. Static threshold approaches are excluded, as prior work has shown them to be both ineffective and inflexible in managing the trade-off between data overhead and accuracy~\cite{bbr}.


\noindent
\underline{\em{BBR Pipe-Full (BBR).}}
The BBR heuristic terminates a speed test once the congestion control algorithm declares the connection ``pipe-full''~\cite{bbr}. We vary the termination threshold by requiring a minimum of $\{1, 2, 3, 5, 7\}$ pipe-full signals before stopping, where larger values reduce premature termination but lower savings. \textcolor{red}{We refer to these configurations as \emph{BBR pipe-$X$}, where $X$ is the required number of pipe-full signals (e.g., \emph{BBR pipe-7, for 7 pipe-full signals till termination}).}

\noindent
\underline{\em{Crucial Interval Sampling (CIS).}}
CIS terminates a test once the throughput distribution stabilizes relative to a reference interval~\cite{fastbts}. We tune the similarity threshold $\beta$ over $\{0.6, 0.8, 0.85, 0.9, 0.95, 1.0\}$, with higher $\beta$ enforcing stricter similarity and improving accuracy at the cost of savings. 

\noindent
\underline{\em{Throughput Stability Heuristic (TSH).}}
This Fast.com-style heuristic terminates a test once throughput variation stabilizes over a sliding window. We adjust the stability threshold over $\{20\%, 30\%, 40\%, 50\%\}$, where smaller thresholds yield more accurate but longer tests. Given that we observe TSH's data savings to be much smaller compared to our previously mentioned baselines, we leave this analysis to Appendix \ref{sec:appendix}.

\smartparagraph{Success metrics.}  
We evaluate each method along two complementary dimensions: accuracy and efficiency. For a given test, let $T$ be the true throughput from a full-length run, $T^{early}$ the throughput estimate at termination, $B$ the bytes transferred in a full run, and $B^{\text{early}}$ the bytes transferred until termination. 

\noindent
\underline{\emph{Data transfer}} is defined as $\frac{B^{\text{early}}}{B}$, where smaller values indicate greater efficiency from early termination. Unless otherwise noted, we report this metric as \emph{cumulative data transferred}, $\frac{\sum_i B_i^{\text{early}}}{\sum_i B_i}$, rather than as per-test averages. This choice reflects the operator’s perspective: what ultimately matters is the aggregate reduction in bandwidth consumption and server-side costs. The cumulative view thus captures the real-world benefit of a stopping policy at scale.
 
\noindent
\underline{\em{Relative error}} is defined as $E_{\text{rel}} = |T - T^{early}|/T$, capturing the accuracy of throughput estimation under early stopping. We use relative error rather than absolute error because speed tests span several orders of magnitude in throughput (e.g., sub-10~Mbps vs. 1~Gbps tiers). An absolute error of 10~Mbps may be negligible in a high-speed tier but catastrophic in a low-speed tier; normalizing by $T$ ensures comparability across tiers. Unless otherwise noted, we report the \emph{median} relative error across tests, which reflects typical accuracy. 


\begin{figure}
    \centering
    \includegraphics[width=0.8\linewidth]{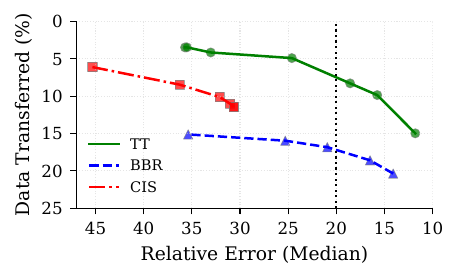}
    \caption{Pareto frontiers of \toolname, BBR, and CIS.}
    \label{fig:pareto_all}
\end{figure}

\subsection{\toolname vs. Existing Strategies}
\label{ssec:eval_global}
We first ask whether \textsc{TurboTest} extends the accuracy--savings frontier beyond existing early termination strategies. Each method exposes a tunable parameter that trades accuracy for efficiency: $\epsilon$ in \textsc{TurboTest}, pipe-full counts in BBR, and similarity threshold $\beta$ in CIS. Because these parameters \textcolor{red} {span  
different operating points, comparing strategies with their default or "fixed" settings alone is uninformative.} Instead, we compare their \emph{Pareto frontiers} in the two-dimensional space of accuracy (median relative error) and efficiency (cumulative data transfer). Figure~\ref{fig:pareto_all} shows these frontiers for \textsc{TurboTest}, BBR, and CIS. The X-axis shows median relative error. The Y-axis reports cumulative data transfer as a percentage of the \textcolor{red}{172.3~TB required if all 1 million tests} in the set ran to completion. The lines denote the frontier for each method with markers indicating parameter settings.

\textsc{TurboTest} dominates across the entire frontier. \textcolor{red}{ At $\epsilon = 20$, it achieves 95.1\% savings (4.9\% transferred) with 24.6\% error, compared to BBR’s maximum of 84.9\% savings (15.1\% transferred) at \textcolor{red}{35}\% error. Across the sweep of $\epsilon$, \textsc{TurboTest} spans 85–96\% savings with 12--36\% error, while BBR covers a similar error range (14–\textcolor{red}{35\%}) but never exceeds 85\% savings. CIS falls in between: its default yields 31.0\% error and 89\% savings, while reducing $\beta$ increases data savings only at the cost of sharply increasing error.}

\textcolor{red}{At the aggressive end of the frontier (approximately 36\% error), \textsc{TurboTest} transfers just 6~TB 
compared to 10.5~TB 
for CIS (45.3\% error) and 26.1~TB 
(35.4\% error) for BBR---a 1.8--4.4$\times$ reduction. At the conservative end ($\epsilon=5$), it transfers 25.8~TB with 11.8\% error, versus 35.1~TB for BBR at 14.1\% error. } 
\textcolor{red}{Detailed metrics for each termination methodology are  enumerated in Table~\ref{tab:data-savings} of Appendix~\ref{sec:appendix}}.  

We now focus on an operationally meaningful target: achieving a median error below 20\% when applying our chosen method on the complete test dataset. From Figure~\ref{fig:pareto_all}, only \textsc{TurboTest} with $\epsilon \leq 15$ and BBR with 
more than five pipe-full signals qualify.  CIS in its configuration with the best accuracy, has a median relative error of 30.7\%, which exceeds our 20\% defined median relative error threshold. Hence, we exclude it from our analysis moving forward. Using the most aggressive parameters (i.e. tending to stop earlier) with median relative error less than 20\%, we observe that \textcolor{red}{\toolname transfers about 14.3~TB, nearly $2.25\times$ less than BBR (32~TB). Put in perspective, this $\sim$92\% savings (8\% transferred) }would reduce M-Lab’s monthly overhead from 12~PB to under 1~PB---a 12$\times$ \textcolor{red}{reduction from the data transferred over fixed-duration tests}. 

\begin{figure}[t]
    \centering
    \begin{subfigure}{.48\linewidth}
      \centering
    \includegraphics[width=\linewidth]{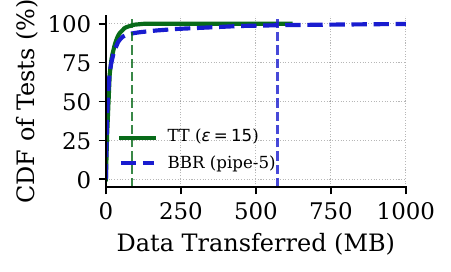}
    \caption{Data transfer distribution}
    \label{fig:cdf_data_transfer}
    \end{subfigure}
    \hfill
    \begin{subfigure}{0.48\linewidth}
    \centering
    \includegraphics[width=1\linewidth]{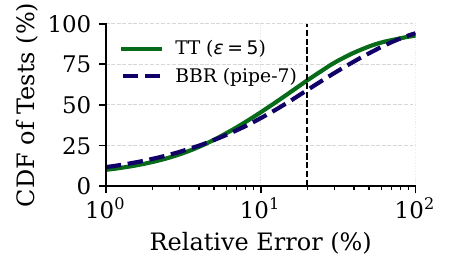}
    \caption{Error distribution}
    \label{fig:global_error_tail}
    \end{subfigure}
    \caption{Distribution of data transfer and relative errors across samples for parameters that satisfy the median error $<20\%$ constraint.}
    \label{fig:inference_time}
\end{figure}





Figure~\ref{fig:cdf_data_transfer} further compares these configurations by plotting the CDF of data transferred per test. While BBR occasionally transfers less data than \textsc{TurboTest} per test, the differences are marginal and disappear once we move to the upper percentiles. Specifically, at the 99th percentile (as visualized by the vertical marking for TT and BBR, respectively) \textcolor{red}{we find that BBR pipe-5 transfers >550 MB for the top 1\% of tests whereas the top 1\% of tests for the most aggressive \toolname model ($\epsilon$=15) transfers only 87 MB (6.3x less)}.  

Figure~\ref{fig:global_error_tail} examines error distributions in these settings when selecting the most conservative (tending to stop late) configurations of  \toolname ($\epsilon$=5) and BBR-pipe 7. Both schemes are heavy-tailed: while both \textsc{TurboTest} and BBR meet the 20\% bound at the median, neither sustains it at higher quantiles. This motivates the need for more adaptive approaches that can trim the error tail, which we explore next in Section~\ref{ssec:deepdive}.




\subsection{Deep Dive: Who Benefits Most?}
\label{ssec:deepdive}
Our previous analysis shows the overall strength of \toolname, but we observe that BBR can outperform \textsc{TurboTest} for certain subsets of tests. This motivates two questions: (i) what characterizes the cases where heuristics perform better, and (ii) which types of tests benefit most from an ML-based approach? To this end, we build on the 20\% error case study from Section~\ref{ssec:eval_global}, decomposing the dataset by speed tiers and RTT bins.  For RTT, we use thresholds at [24, 52, 115, 234]~ms, which approximately correspond to the 25$^{th}$, 50$^{th}$, 75$^{th}$, and 90$^{th}$ percentiles of our dataset.  This decomposition enables a systematic comparison of strategies across heterogeneous network conditions and reveals where learning-based methods deliver the largest advantages.

\begin{figure}[t]
    \centering
    \includegraphics[width=1\linewidth]{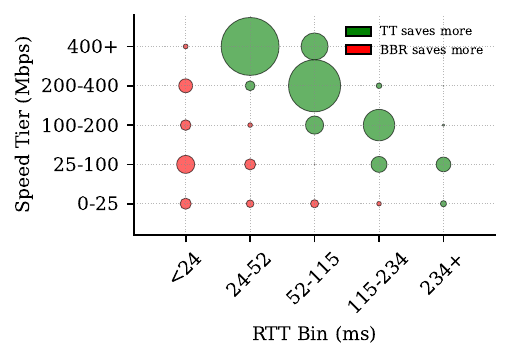}
    \caption{Delta in data transfer between \toolname and BBR across all speed-tier × RTT groupings. Point size indicates the magnitude of the difference in data transferred. Green denotes cases where \toolname transfers less, while red denotes cases where BBR transfers less.}
    \label{fig:deepdive}
    \vspace*{-0.08in}
\end{figure}

\begin{figure*}[t]
    \centering
    \begin{subfigure}{0.31\linewidth}
       \centering
    \includegraphics[width=1\linewidth]{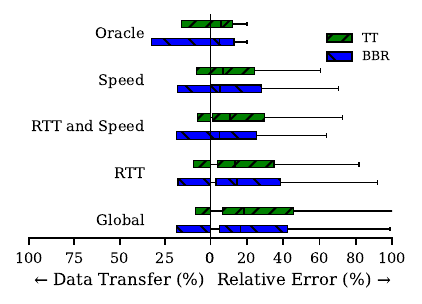}
    \caption{Comparison of cumulative data transferred (left) and relative error (right) across different strategies }
    \label{fig:adaptive_composite}
    \label{fig:bbr_dist}
    \end{subfigure}
    \hfill
    \begin{subfigure}{0.31\linewidth}
      \centering
    \includegraphics[width=1\linewidth]{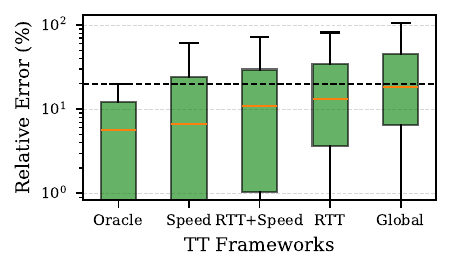}
    \caption{Distribution of relative error for different \toolname frameworks (20\%  relative error line is marked for clarity)}
    \label{fig:cdf_frameworks}
    \end{subfigure} \hfill
     \begin{subfigure}{0.31\linewidth}
          \centering
    \includegraphics[width=1\linewidth]{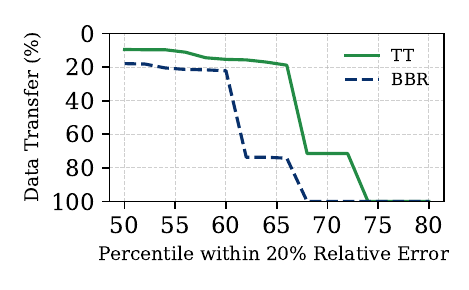}
    \caption{Data transfer across RTT-aware termination strategies at varying relative error (20\%) percentiles}
    \label{fig:data_savings_percentile}
    \end{subfigure}
    \caption{Adaptive parameterization strategies. The label TT refers to \textsc{TurboTest}.
}
    \label{fig:adaptive parameterization}
\end{figure*}
Figure~\ref{fig:deepdive} summarizes the results using a matrix visualization across all 25 speed tier $\times$ RTT bin combinations. Each cell reports the relative advantage of \toolname versus BBR when both are tuned to their most aggressive parameter that still satisfies the median error $<$20\% constraint identified earlier. The size of each bubble is proportional to the absolute difference in cumulative data transferred, while color encodes the winner: green indicates that \textsc{TurboTest} transfers less data, red indicates BBR transfers less. We observe that \toolname outperforms BBR in high-speed tiers, which explains its overall superior data savings, as these tiers contribute the most to total data transfer (see Figure~\ref{fig:distribution_skew}). We attribute these gains to the use of a balanced training dataset, which provides better opportunities to learn termination behavior for relatively scarce but critical high-speed test cases. Moreover, \toolname also performs better for tests with higher RTTs, underscoring ML’s superior ability to make effective use of fewer samples in challenging high-latency conditions.

\smartparagraph{Takeaway.}
Overall, this deep dive demonstrates that \textsc{TurboTest} not only dominates heuristics in aggregate but also offers the best accuracy-savings trade-off across most operational scenarios. Its largest gains occur in precisely those tests---high throughput, moderate RTT---that contribute most to aggregate data volume.

\vspace*{-0.1in}
\subsection{Adaptive Parameterization}
\label{sec:adaptive}
\smartparagraph{Grouping strategies for adaptive parameterization.}
Thus far, we have considered global settings or coarse group-wise tuning (by speed tier or RTT bin). A practical question remains: \emph{how should parameters be selected to suppress tail errors while retaining efficiency?} We formalize this as a constrained selection problem and evaluate five strategies, all under the same rule: within each strategy’s grouping scope, we sweep each method’s control knob (\textsc{TurboTest} $\epsilon$ and BBR pipe-full count) and pick the \emph{most aggressive} setting that \emph{keeps the group’s median relative error below $20\%$}; if no setting satisfies the constraint for a group, that group does not terminate early. Concretely:
(1) \emph{Global strategy}: one group (the entire test set); a single parameter is applied to all tests (as in Section~\ref{ssec:deepdive}).
(2) \emph{Speed-only strategy}: groups are throughput tiers; we select one parameter per tier.
(3) \emph{RTT-only strategy}: groups are RTT bins; we select one parameter per bin.
(4) \emph{RTT+Speed strategy}: groups are all $(\text{tier},\text{RTT})$ pairs; we select one parameter per pair.
(5) \emph{Oracle strategy}: groups degenerate to single tests; for each test we choose the most aggressive setting that would keep that test’s relative error $\le 20\%$, otherwise we do not terminate it. The Oracle represents a theoretical upper bound on what grouping can achieve.  Detailed information on the specific method chosen for each speed/RTT class is in Appendix~\ref{sec:appendix}.

Figure~\ref{fig:adaptive_composite} summarizes the results using a composite plot: for each strategy, we show cumulative data transferred (bars on left) alongside the distribution of relative errors (box plots on right) for \textsc{TurboTest} and BBR.  Several observations emerge. First, both Speed-only and RTT-only strategies substantially reduce tail errors compared to the Global strategy, while RTT+Speed offers even greater flexibility. This is further validated in Figure~\ref{fig:cdf_frameworks}, which highlights how adaptive parametrization in the case of \toolname reduces the median relative error over the simplistic "global" strategy. Unsurprisingly, the Oracle strategy provides the best possible trade-off, reflecting an unattainable upper bound in realizing low relative error for the tail. 
Second, across all comparisons, \textsc{TurboTest} dominates BBR by transmitting $2\times$ less data while maintaining comparable tail errors under equivalent accuracy constraints. 

The practical implication lies in the choice between Speed-only and RTT-only grouping. While Speed-only adaptation achieves improvements, it is difficult to deploy because the throughput tier cannot be reliably inferred in the first few hundred milliseconds of a test. RTT-only grouping, by contrast, is practical: RTT can be measured immediately at runtime and provides a strong, deployable basis for adaptation. Thus, RTT-aware parameterization emerges as the most effective middle ground between the simplicity of global tuning and the theoretical optimality of the Oracle.



\smartparagraph{Taming the tails.}
Finally, we evaluate how well different schemes can contain \emph{tail errors}, i.e., performance beyond the median, under increasingly strict accuracy constraints. 
Figure~\ref{fig:data_savings_percentile} compares our RTT-aware \textsc{TurboTest} framework with BBR when progressively tightening the error requirement from the median to higher quantiles. The evaluation procedure is as follows: for each method, we first ensure that the average (median) test is within a $20\%$ relative error bound. We then select the best possible configuration for data savings under this constraint and measure how savings degrade as the percentile error requirement becomes stricter. The results show a clear separation. At the median ($50^{\text{th}}$ percentile), both methods can be tuned to satisfy the $20\%$ constraint, but \textsc{TurboTest} consistently achieves higher data savings for the same error bound. \textcolor{red}{ When looking at the raw bytes transferred for this region, this translates to $2 \times$ less data transferred than BBR. As we move to the $62^{\text{th}}$-- $66^{\text{th}}$ percentiles, \textsc{TurboTest}'s RTT-aware parameterization sustains at less than 20\% data transferred, whereas BBR's increases to 60\% or more. More specifically, even in regimes where operators may enforce higher accuracy across a larger fraction of tests, \textsc{TurboTest} transfers between $1.4\times$ and $4.7\times$ less data than BBR.} This demonstrates that adaptive use of richer features allows \textsc{TurboTest} to preserve efficiency while curbing errors in the bulk of the distribution. 

\begin{figure}[t]
    \centering
    \begin{subfigure}{0.49\linewidth}
        \centering
        \includegraphics[width=\linewidth]{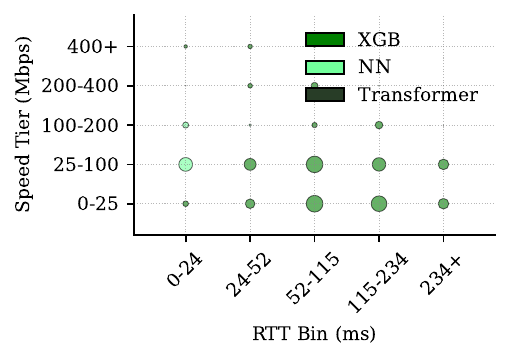}
        \caption{Model architecture
        }
        \label{fig:regressor_variants}
    \end{subfigure}
    \begin{subfigure}{0.49\linewidth}
        \centering
        \includegraphics[width=\linewidth]{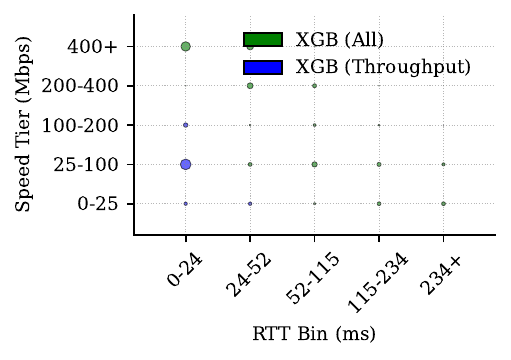}
        \caption{Features
        }
        \label{fig:xgb_variants}
    \end{subfigure}
    
    \caption{Delta in data transfer across regressors. 
    Point size indicates the magnitude of the difference in data transferred, while color indicates the best performing method. }
    \label{fig:regressor_ablation}
\end{figure}

However, the figure also highlights the fundamental limits of early termination. \textcolor{red}{ At the $74^{\text{th}}$ percentile,} no method---including \textsc{TurboTest}---is able to achieve safe early termination, with data savings collapsing to $0\%$ (100\% transferred). Roughly one-quarter of tests remain resistant to early termination without inflating prediction error. These hard cases are typically low-throughput flows with high RTT, where variability persists for most of the test duration. This finding underscores both the progress and the boundaries of current approaches: while adaptive ML-based parameterization can dramatically expand the accuracy-savings frontier, a significant fraction of speed tests remain inherently difficult to truncate early, reflecting the complexity of network dynamics in challenging environments.

\smartparagraph{Takeaway.}
Adaptive parameterization, particularly RTT-aware strategies, provides operators with a practical mechanism to curb tail errors while retaining most of the efficiency benefits of \textsc{TurboTest}. While oracle-level adaptivity is infeasible, RTT-aware grouping demonstrates that careful parameterization can deliver robust performance at scale.

\begin{figure}[t]
    \centering
    \includegraphics[width=1\linewidth]{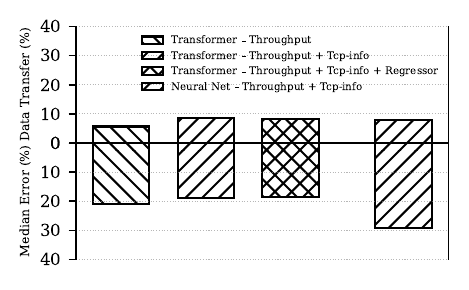}
    \caption{Classifier performance under a fixed XGBoost (TCPInfo) regressor.}
    \label{fig:classifier_varients}
\end{figure}

\subsection{Ablation Study}
\label{sec:ablation}
Next, we validate that \textsc{TurboTest}'s dual-stage ML design is justified. We systematically swap regressors (XGBoost, Neural Net, Transformer), classifiers, and feature subsets (throughput-only, TCPInfo-only, hybrids), as summarized in Figures~\ref{fig:regressor_ablation} and \ref{fig:classifier_varients}. We exclude XGBoost from the classifier analysis because its architecture is not  easily amenable to training with full length sequences.  This results from   issues  fitting our entire classification dataset into memory at once.\looseness+1 

\smartparagraph{Regressor (stage~1).}
The results are consistent: for stage~1 regression, XGBoost offers the strongest predictive accuracy across throughput and TCPInfo features, while stage~2 classification with a Transformer yields the best stopping decisions and savings-accuracy balance. 
In Figure~\ref{fig:regressor_ablation},  the ideal stopping point for each regressor configuration is found, i.e. the earliest possible timestamps with a relative error less than or equal to 20\%. We then visualize the delta in the data transferred for all test classes (speed tier+RTT) through this constraint. Figure~\ref{fig:regressor_variants} illustrates that, in a majority of classes, with the exception of the 0-24ms RTT -- 25-200Mbps tests, XGboost outperforms the other regressors with higher data savings due to typically earlier stopping points. This is especially evident in the median latency low throughput tiers, through the higher deltas. This indicates that XGboost is able to capture dynamics that other models miss. 
Recent research has shown that Gradient boosted decision trees excel in handling heavy-tailed distributions and skews, which are inherent in network data~\cite{mcelfresh2024neuralnetsoutperformboosted}.  Figure~\ref{fig:xgb_variants} shows that adding in TCP-info features only marginally helps reduce the ideal stopping point/time, with the delta between the throughput and TCP-info variants remaining small.

\begin{figure}[t]
    \centering
    \includegraphics[width=0.8\linewidth]{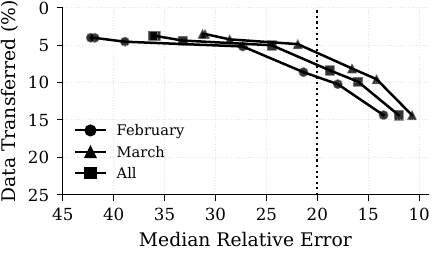}
    \caption{Pareto frontiers for the new distribution in February and March.}
    \label{fig:temporal-frontiers}
\end{figure}

\smartparagraph{Classifier (stage~2).}
Figure~\ref{fig:classifier_varients} compares transformer-based classifiers at $\epsilon=15$. All three variants achieve broadly similar performance, with the throughput-only model showing a modestly higher median relative error (21\%) compared to those augmented with TCP-info or regressor outputs. The addition of TCP-info slightly improves stopping accuracy, though the incremental gains over throughput alone are relatively small. By contrast, our end-to-end neural network variant transfers less data but exhibits substantially higher error (29\%), underscoring that it fails to reliably capture stopping dynamics. TCP-info features offer only marginal improvements over throughput alone, so the main benefit comes from using a Transformer classifier itself rather than the exact feature mix.

\smartparagraph{Takeaway.}
These ablations demonstrate that both stages are essential. 
XGBoost in Stage~1 provides robust regression accuracy, especially in skewed and heavy-tailed network conditions, while the Transformer in Stage~2 delivers strong stopping decisions. 
Using more elaborate feature sets (e.g., TCPInfo) offers only marginal gains, while applying the same model type across both stages (e.g., Transformer-only) reduces performance. 
Together, these results confirm that \toolname's two-stage design is well-founded and critical for achieving its accuracy--savings frontier.

\subsection{Robustness and Overheads}
\label{ssec:robustness-overheads}

\smartparagraph{Robustness to concept drift.}
To assess robustness under temporal drift, we evaluate all $\epsilon$ configurations of \textsc{TurboTest} on the \emph{robustness set}---133k speed tests collected in February and March 2025---while the model was trained exclusively on April 2024--January 2025 data. Figure~\ref{fig:temporal-frontiers} plots Pareto frontiers for February and March alongside the frontier for the 2024--2025 training period, allowing us to directly visualize how accuracy--savings trade-offs evolve over time. Overall, we observe only mild drift: median relative error shifts by less than 2\% across the full robustness set. However, February exhibits larger deviations than March, with roughly 4\% higher median error at $\epsilon=15$. Inspecting the data reveals that the February set contains more low-throughput, high-RTT tests, concentrated in the 90$^{th}$ percentile RTT bin. This suggests that \textsc{TurboTest} remains consistent in which regimes it performs well (stable, higher-throughput connections) and where it struggles (sparse, high-latency flows), underscoring the need for periodic retraining to maintain accuracy as the distribution of test conditions evolves.

\smartparagraph{Training overhead.}
\textcolor{red}{
We measure \textsc{TurboTest}’s offline training cost to evaluate training across all $\epsilon$ configurations. Stage~1 is $\epsilon$-independent (fit XGBoost once on the full training set), while Stage~2 trains a transformer (classifier) per $\epsilon$. On a 4$\times$A100 node (256~GB RAM) with a 64-core AMD CPU, Stage~1 on 800k tests takes 14 minutes and Stage~2 (5 epochs) takes $\sim$50 minutes, totaling 64 minutes per $\epsilon$ (5.8 hours for seven). Since training is offline and can be parallelized, this can make the effective training time to 1.06 hours, making full retraining practical at deployment scale.}

\smartparagraph{Runtime overhead.}  
A key question for deployment is whether \textsc{TurboTest} can operate in real time without incurring prohibitive inference costs. In particular, the use of a Transformer-based classifier for stopping decisions may appear impractical for online measurement pipelines. To assess feasibility, we measure the wall-clock inference time of both stages of the pipeline: the XGBoost regressor (Stage~1) and the Transformer classifier (Stage~2). Our measurements report the latency from the arrival of a new TCPInfo snapshot to the model’s output, excluding TCPInfo preprocessing time since that step is implementation-dependent and typically amortized.  

For Stage~1, we simulate speed test execution by feeding TCPInfo snapshots in 500~ms increments and running inference across varying batch sizes that mimic the workload of M-Lab measurement servers (from a handful of concurrent tests up to nearly 1{,}000). The regressor consistently produces predictions within 10~ms, averaging 6.3~ms, with only mild increases as batch size grows. For Stage~2, we run the Transformer classifier in tandem with Stage~1 after warm-up. Once in steady state, classification decisions are produced within 14~ms on average, with stable latency across batch sizes; occasional outliers (up to 133~ms) stem from scheduling delays rather than the model itself.  

\smartparagraph{Takeaway.} Overall, these results demonstrate that \textsc{TurboTest} can comfortably meet real-time constraints: both prediction and stopping decisions are returned an order of magnitude faster than the 500~ms decision interval. While further runtime optimizations could reduce latency even more, the current design already establishes the feasibility of deploying \textsc{TurboTest} in production environments.

\section{Related Work}
\label{sec:related}
\smartparagraph{Lightweight speed tests.} Various speed testing approaches have evolved that aim to balance accuracy and efficiency. Early work relied on UDP probing with crafted packets to estimate available bandwidth~\cite{pathload, pathchirp, igi, assolo}. For instance,  Pathload uses periodic packet trains~\cite{pathload}, while Pathchirp introduces exponentially spaced bursts~\cite{pathchirp}. These methods face challenges from fluid traffic assumptions, interrupt coalescing, differential treatment of UDP traffic, and timestamp inaccuracies~\cite{lao2006probe}, challenges that worsen in high-speed networks. \textcolor{red}{Orthogonal to active speed testing, prior work has explored passive  bandwidth estimation by piggybacking on background user traffic (e.g., pre-staged ads)~\cite{mohammed2023repurpose, mohammed2023harmful} or synthesizing estimates from network telemetry~\cite{9209675}.} 

Recent efforts in active measurement include FastBTS, which modifies the probing and termination for TCP-based tests. Gill et al.~\cite{bbr} design a heuristic that leverages transport signals for early termination. The most closely related work comes from Maier et al.~\cite{maier2019reducing} and Arifuzzaman et al.~\cite{arifuzzaman2022swift}, who apply ML for early termination. Maier et al.~\cite{maier2019reducing} restrict the problem to a fixed-point binary decision, while Arifuzzaman et al.~\cite{arifuzzaman2022swift} consider both regression and classification, but with fixed-input models that rely solely on throughput signals and are evaluated only in emulated conditions. In contrast, we propose a general, data-driven approach extensively validated on real-world network data.

\smartparagraph{Optimal stopping problem.} Optimal stopping is a classic problem with applications in finance, industry, and healthcare. Traditional solutions are based on modeling, leveraging stochastic processes~\cite{peskir2006optimal} and control theory~\cite{tsitsiklis2002optimal}. However, due to the non-stationary nature of network conditions, data-driven approaches are more relevant. The closest ML-based line of work is early time-series classification~\cite{gupta2020approaches, bondu2022open}, which seeks to classify a sequence as early as possible. Techniques  include combining a predictor (a classifier in this case) with heuristic termination rules (e.g., confidence thresholds or stability checks)~\cite{ghalwash2014utilizing}, identifying discriminative shapelets~\cite{yan2020extracting}, or incorporating the cost of delay into the optimization objective~\cite{dachraoui2015early}. A key distinction is that our setting involves the regression task, where confidence-based stopping is less natural and more challenging.

\smartparagraph{Improving speed tests.} Prior studies have analyzed the impact of factors such as test protocol, network conditions, and server infrastructure on speed test accuracy~\cite{macmillan2022best, macmillan2023comparative, feamster2020measuring, zhang2024empirical}. Another line of work contextualizes these factors into the crowdsourced data before using it, especially for policy purposes~\cite{paul2022importance, bauer2010understanding, clark_measurement_2021}. These efforts are complementary to ours, which focuses on improving the efficiency of tests.
\vspace{-9pt} 
\section{Future Work and Conclusion}
\label{sec:discussion}

Our evaluation on \textcolor{red}{1 million} M-Lab NDT tests  shows that \textsc{TurboTest} achieves 1.8-$4.4\times$ higher data savings than M-Lab’s BBR-based approach while reducing median error, shifting the Pareto frontier outward and approaching oracle bounds. These results highlight the limits of existing heuristics—static thresholds, BBR pipe-full signals, and throughput stability rules—that capture only a narrow slice of the accuracy–savings trade-off. By decoupling prediction from termination, leveraging richer transport-level features, and exposing a single tunable parameter $\epsilon$, \textsc{TurboTest} delivers accurate, efficient, and robust termination policies suitable for deployment at scale.

Several avenues exist for future work. Alternate loss functions could mitigate mean-squared error’s bias toward high-speed tiers, and more advanced architectures (e.g., autoregressive transformers) may improve accuracy, especially at the tail. For termination, reinforcement learning could be used to learn policies without explicit labels but would likely require substantially larger datasets. \textcolor{red}{ Time-series models geared towards unsupervised learning may be explored to further improve tail performance \cite{guthula2025netbursteventcentricforecastingbursty}. }Beyond modeling, the methodology generalizes naturally: while we focused on download tests due to their higher data overhead, the same approach applies to upload tests. Similarly, while our evaluation used NDT because of its public data, the framework is extensible to other speed test applications, including multi-connection tests or tools that report throughput using different aggregation rules. \textcolor{red}{Future work may include comparing \textsc{TurboTest} variants in controlled environments~\cite{daneshamooz2025netreplicaprogrammablesubstratelastmile}.} Importantly, our approach is agnostic to what a speed test should report -- a complementary question outside the scope of this work. 

This work provides the first principled learning formulation for speed test optimization, opening opportunities for adaptive measurement frameworks that generalize across heterogeneous networks while reducing the resource cost of a critical public service. In short, \textsc{TurboTest} shows that less can indeed be enough.

\section{Acknowledgment}
\textcolor{red}{We thank the anonymous reviewers and our shepherd, Zahaib Akhtar, for their constructive feedback and suggestions. We also acknowledge Sylee Beltiukov for their contributions to architecting \textsc{TurboTest}'s production-ready implementation.} This work was supported by the National Science Foundation (Graduate Research Fellowship Grant No. 2139319, CAREER Award No. 2443777, and CNS Award No. 2323229), research gifts from Cisco and Google, and by the Anusandhan National Research Foundation (ANRF) under Grant No. ANRF-ECRG-2024-005839. This research used resources of the National Energy Research Scientific Computing Center, a DOE Office of Science User Facility supported by the Office of Science of the U.S. Department of Energy under Contract No. DE-AC02-05CH11231 using NERSC award NERSC DDR-ERCAP0029768.

\label{lastpage}

\microtypesetup{protrusion=false}
\end{sloppypar}

\bibliographystyle{plain}
\bibliography{references}

\appendix
\clearpage
\section{Appendix} \label{sec:appendix}

\subsection{TurboTest vs Existing Strategies}
\begin{table}[H]
\centering
\footnotesize
\setlength{\tabcolsep}{4pt}
\renewcommand{\arraystretch}{1.05}
\resizebox{\columnwidth}{!}{%
\begin{tabular}{lcc}
\toprule
\textbf{Method} &
\textbf{Data Transferred (TB / \%)} &
\textbf{Median Rel. Err. (\%)} \\
\midrule
TT ($\epsilon = 5$)  & 25.8 / 15.0\% & 11.8 \\
TT ($\epsilon = 10$) & 17.0 / 9.9\%  & 15.8 \\
TT ($\epsilon = 15$) & 14.3 / 8.3\%  & 18.6 \\
TT ($\epsilon = 20$) & 8.5  / 4.9\%  & 24.6 \\
TT ($\epsilon = 25$) & 6.0  / 3.5\%  & 35.7 \\
TT ($\epsilon = 30$) & 6.0  / 3.5\%  & 35.5 \\
TT ($\epsilon = 35$) & 7.2  / 4.2\%  & 33.0 \\
\midrule
BBR-pipe 1 & 26.1 / 15.1\% & 35.4 \\
BBR-pipe 2 & 27.5 / 16.0\% & 25.3 \\
BBR-pipe 3 & 29.0 / 16.8\% & 20.9 \\
BBR-pipe 5 & 32.0 / 18.6\% & 16.5 \\
BBR-pipe 7 & 35.1 / 20.4\% & 14.1 \\
\midrule
CIS $\beta = 0.6$  & 10.5 / 6.1\%  & 45.3 \\
CIS $\beta = 0.80$ & 14.6 / 8.5\%  & 36.3 \\
CIS $\beta = 0.85$ & 17.4 / 10.1\% & 32.1 \\
CIS $\beta = 0.90$ & 19.0 / 11.0\% & 31.0 \\
CIS $\beta = 0.95$ & 19.7 / 11.4\% & 30.7 \\
CIS $\beta = 1.0$  & 19.7 / 11.5\% & 30.7 \\
\midrule
No Termination & 172.3 / 100.0\% & -- \\
\bottomrule
\end{tabular}%
}

\vspace{0.3em}
\caption{Median Relative Error (\%) and Data transferred (in Terabytes and as a fraction of the data transferred with no early termination strategy) across termination methodologies for the 1 million M-lab NDT Tests.}
\label{tab:data-savings}
\end{table}

\subsection{Analysis of Throughput Stability Heuristic (TSH)}\label{a1}

We apply TSH on our test dataset of 1 milion samples and calculate metrics such as Median Relative Error and Data Transfer as visualized in Table \ref{tab:tsh_results}. As one can see, by increasing the stability threshold, the amount of data being transferred decreases at the cost of relative error. The least amount of data (greatest data savings) is transferred with the stability threshold value of 50 at 34.9\%. However, TT ($\epsilon$=5) for comparison, which is the most conservative TT configuration, has a data transfer value which is much lower at 15\%. One can conclude, that TSH excels when the operator's primary concern is prediction accuracy, with data savings being a secondary factor. 

\begin{table}[H]
\centering
\resizebox{\columnwidth}{!}{%
\begin{tabular}{cccc}
\toprule
\textbf{Stability Threshold} & 
\makecell{\textbf{Median Relative}\\\textbf{Error (\%)}} & 
\makecell{\textbf{Data}\\\textbf{Transfer (\%)}} & 
\makecell{\textbf{Data}\\\textbf{Transferred (TB)}} \\
\midrule
20 & 0 & 70.7 & 121.8 \\
30 & 0 & 52.9 & 91.1 \\
40 & 0.91 & 41.7 & 71.9 \\
50 & 2.73 & 34.9 & 60.3 \\

\bottomrule
\end{tabular}}
\caption{Median relative error and data transfer for TSH configurations.}
\label{tab:tsh_results}
\end{table}

\subsection{Configurations for Adaptive Parameterization}\label{a2}

Tables \ref{tab:speed_tier} and \ref{tab:rtt-bin} report the best configurations of \textsc{TurboTest}, BBR, and CIS when stratified by speed tier and RTT. For each subset of tests (e.g., 0–25 Mbps low-throughput tests), we select the parameter setting for each method that yields the maximum data savings while keeping the overall median relative error below 20\%. This customization allows us to explore scenarios in which different methods are preferable depending on the test type. An empty cell indicates that no parameter setting for that method satisfied the 20\% median relative error threshold. As shown, all methods struggle to balance relative error and data transfer in the 0–25 Mbps (low-throughput) tier, whereas CIS also performs poorly in several higher-throughput tiers. Similarly, we also notice the pattern of all methods struggling to terminate in high RTT scenarios (>234ms).

Table \ref{tab:speed-rtt-bin} presents the \textsc{TurboTest} configurations after jointly stratifying the dataset by speed tier and RTT. Some bins are marked as having "No tests" because there are few or no tests in these relatively rare categories. This pattern reflects the common empirical tendency for higher-throughput tests to also exhibit lower latency. For brevity, we omit the analogous tables for BBR and CIS.

\begin{table}[H]
\centering
\scriptsize
\setlength{\tabcolsep}{4pt}
\caption{Best configuration for speed-tier strategy.}
\label{tab:speed_tier}
\begin{tabular}{lccccc}
\toprule
        & 0--25 & 25--100 & 100--200 & 200--400 & 400+ \\
\midrule
ML  & --- & ($\epsilon{=}10$) & ($\epsilon{=}30$) & ($\epsilon{=}30$) & ($\epsilon{=}30$) \\
BBR & --- & pipe-3 & pipe-2 & pipe-1 & pipe-2 \\
CIS & --- & --- & --- & --- & $\beta{=}0.80$ \\
\bottomrule
\end{tabular}
\end{table}

\begin{table}[H]
\centering
\scriptsize
\setlength{\tabcolsep}{4pt}
\caption{Best  configuration for RTT strategy.}
\label{tab:rtt-bin}
\begin{tabular}{lccccc}
\toprule
Method & $<$24ms & 24--52ms & 52--115ms & 115--234ms & 234ms+ \\
\midrule
TT  & $\epsilon{=}15$ & $\epsilon{=}15$ & $\epsilon{=}15$ & $\epsilon{=}5$  & --- \\
BBR & pipe-5          & pipe-2          & pipe-3          & pipe-7          & --- \\
CIS & $\beta{=}0.80$  & $\beta{=}0.90$  & ---             & ---             & --- \\
\bottomrule
\end{tabular}
\end{table}


\begin{table}[H]
\centering
\scriptsize
\setlength{\tabcolsep}{4pt}
\caption{Best TT configuration for RTT+speed strategy.}
\label{tab:speed-rtt-bin}
\begin{tabular}{lccccc}
\toprule
Speed (Mbps) & $<$24\,ms & 24--52\,ms & 52--115\,ms & 115--234\,ms & $\geq$234\,ms \\
\midrule
0--25   & --- & $\varepsilon{=}5$  & $\varepsilon{=}5$  & --- & --- \\
25--100 & $\varepsilon{=}5$  & $\varepsilon{=}15$ & $\varepsilon{=}10$ & $\varepsilon{=}10$ & $\varepsilon{=}5$ \\
100--200& $\varepsilon{=}35$ & $\varepsilon{=}30$ & $\varepsilon{=}30$ & $\varepsilon{=}25$ & --- \\
200--400& $\varepsilon{=}25$ & $\varepsilon{=}30$ & $\varepsilon{=}30$ & $\varepsilon{=}35$ & $\varepsilon{=}5$ \\
400+    & $\varepsilon{=}25$ & $\varepsilon{=}30$ & $\varepsilon{=}30$ & No tests & $\varepsilon{=}10$ \\
\bottomrule
\end{tabular}
\end{table}

\vfill
\pagebreak

\end{document}